\documentclass[a4paper,12pt]{article}

\usepackage[english]{babel}
\usepackage[a4paper,tmargin=3truecm,bmargin=3truecm,rmargin=2.5truecm,
lmargin=2.5truecm,twoside,verbose=true]{geometry}

\usepackage{cancel,graphicx}
\usepackage[dvips]{hyperref}

\usepackage{amsmath,amssymb}

\numberwithin{equation}{section}

\allowdisplaybreaks[1]

\renewenvironment{thebibliography}[1]
         {\section*{References}\frenchspacing\small
          \begin{list}{[\arabic{enumi}]}
         {\usecounter{enumi}\parsep=2pt\topsep 0pt
         \settowidth{\labelwidth}{[#1]}
         \leftmargin=\labelwidth\advance\leftmargin\labelsep
         \rightmargin=0pt\itemsep=1pt\sloppy}}{\end{list}}

\title{Noncommutative Induced Gauge Theory\footnote{Work 
supported by ANR grant NT05-3-43374 ``GenoPhy''.}}
\author{Axel de Goursac$^{a}$, Jean-Christophe Wallet$^a$, 
Raimar Wulkenhaar$^b$}
\date{}

\begin{document}

\maketitle
\vspace*{-1cm}
\begin{center}
\textit{$^a$Laboratoire de Physique Th\'eorique, B\^at.\ 210\\
    Universit\'e Paris XI,  F-91405 Orsay Cedex, France\\
    e-mail: \texttt{axelmg@melix.net}, 
\texttt{jean-christophe.wallet@th.u-psud.fr}}\\[1ex]
\textit{$^b$Mathematisches Institut der Westf\"alischen
  Wilhelms-Universit\"at \\Einsteinstra\ss{}e 62, D-48149 M\"unster,
  Germany \\
    e-mail: \texttt{raimar@math.uni-muenster.de}}\\
\end{center}%

\vskip 2cm

\begin{abstract}
  We consider an external gauge potential minimally coupled to a
  renormalisable scalar theory on 4-dimensional Moyal space and
  compute in position space the one-loop Yang-Mills-type effective
  theory generated from the integration over the scalar field.  We
  find that the gauge invariant effective action involves, beyond the
  expected noncommutative version of the pure Yang-Mills action,
  additional terms that may be interpreted as the gauge theory
  counterpart of the harmonic oscillator term, which for the
  noncommutative $\varphi^4$-theory on Moyal space ensures
  renormalisability. The expression of a possible candidate for a
  renormalisable action for a gauge theory defined on Moyal space is
  conjectured and discussed.
\end{abstract}%

\pagebreak

\section{Introduction.}

In the past few years, an intense activity has been devoted to the
study of various classes of field theories defined on Moyal spaces
(see e.g.\ \cite{Douglas:2001ba,Wulkenhaar:2006si}). These prototypes
of noncommutative field theories involve numerous features stemming
from noncommutative geometry \cite{Grossmann:1968,CONNES,CM} and are thus interesting
in themselves. This interest was further increased by the observation
that similar noncommutative field theories seem to emerge rather
naturally from limiting regimes of string theory and matrix theory in
magnetic backgrounds \cite{Seiberg:1999vs,Schomerus:1999ug}. See also
\cite{Witten:1985cc,Connes:1997cr} for connections between
noncommutative geometry and string theory. Recall that in
noncommutative geometry the commutative algebras of functions defined
on differentiable manifolds (roughly speaking the coordinates spaces)
are replaced by associative but noncommutative algebras further
interpreted as algebras of functions on ``noncommutative spaces''.
Within this algebraic framework, natural noncommutative analogues of
the main geometrical objects usually involved in field theories can be
algebraically defined (such as connections, curvatures, vector
bundles) so that the construction of various noncommutative analogues
of field theories can be undertaken (see e.g.\ \cite{Gayral:2006wu}).
The starting relevant configuration spaces for the noncommutative
field theories are modules over the associative algebras which are
naturally viewed as noncommutative analogues for the set of sections
of vector bundles.  One example of associative algebra among many
others is provided by the associative Moyal algebras
\cite{Gracia-Bondia:1987kw,Varilly:1988jk} therefore playing the role
of ``noncommutative Moyal spaces''.\par

The simplest generalisations of scalar theories to Moyal space were
shown to suffer from the so called UV/IR-mixing
\cite{Minwalla:1999px,Chepelev:1999tt}, a phenomenon that makes the
renormalisability very unlikely. Basically, the UV/IR-mixing results
from the existence of potentially dangerous non-planar diagrams which,
albeit UV finite, become singular at exceptional (low) external
momenta. This triggers the occurrence of UV divergences in higher
order diagrams in which they are involved as subdiagrams. This signals
that UV and IR scales are related in a non-trivial way which should in
principle invalidate a Wilson-type renormalisation scheme
\cite{Wilson:1973jj,Polchinski:1983gv}. An appealing solution to the
UV/IR-mixing has been recently proposed by Grosse and Wulkenhaar
\cite{Grosse:2004yu,Grosse:2003aj} within the noncommutative
$\varphi^4$ model on the 4-dimensional Moyal space where $\varphi$ is
real-valued. They showed that the UV/IR-mixing can be suppressed by
supplementing the initial action with a harmonic oscillator term
leading to a renormalisable noncommutative quantum field theory. The
initial proof \cite{Grosse:2004yu} was performed within the
matrix-base formalism, roughly speaking a basis for the (Schwarz
class) functions for which the associative product of the Moyal
algebra is a simple matrix product.  This cumbersome proof was
simplified through a reformulation into the (position) $x$-space
formalism in \cite{Gurau:2005gd} which exhibits some advantages compared to
the matrix-base formulation. For instance, the propagator in $x$-space
can be explicitly computed (as a Mehler kernel \cite{Simon,Gurau:2005qm})
and {\it{actually}} used in calculations.  Besides, it makes easier
the comparison of the renormalisation group for noncommutative
theories and their commutative counterpart.\par

Other renormalisable noncommutative matter field theories on Moyal
spaces have been obtained. One is the complex-valued scalar theory
studied in \cite{Gurau:2005gd} which can be viewed as a modified
version of the LSZ model \cite{Langmann:2003if,Langmann:2003cg} (the
scalar theory in \cite{Grosse:2003nw} is super-renormalisable). Note
that interesting solvable noncommutative scalar field theories have
also been considered in
\cite{Grosse:2005ig,Grosse:2006qv,Grosse:2006tc}. As far as fermionic
theories are concerned, a Moyal space version of the Gross-Neveu model
\cite{Gross:1974jv} (see also \cite{Mitter:1974cy,Kopper:1993mj}),
called the orientable noncommutative Gross-Neveu model, has been
recently shown to be renormalisable to all orders
\cite{Vignes-Tourneret:2006nb,Vignes-Tourneret:2006xa} (see also
\cite{Lakhoua:2007ra}). It is worth mentioning that this
noncommutative field theory still exhibits some UV/IR-mixing, even in
the presence of the fermionic version of the harmonic oscillator
quadratic term introduced in \cite{Grosse:2004yu}, which however does
not prevent the theory from being renormalisable. Note that in
\cite{Akhmedov:2001fd} (see also \cite{Akhmedov:2000uz}) the large-$N$
limit of the noncommutative Gross-Neveu model, however with a
restricted interaction, has been studied; renormalisability is shown
at this limit together with asymptotic freedom. One should keep in
mind that the fact that the orientable Gross-Neveu model is
renormalisable in spite of some remaining UV/IR-mixing
\cite{Vignes-Tourneret:2006nb,Vignes-Tourneret:2006xa} indicates that
further investigations are needed to actually clarify the role of
various generalisations of the above-mentioned harmonic oscillator
term, of the related covariance under the Langmann-Szabo duality
\cite{Langmann:2002cc} and of their impact in the control of the
UV/IR-mixing and renormalisability.\par

So far, the construction of a renormalisable gauge theory on
noncommutative Moyal spaces remains still unsolved. The naive
noncommutative extension of the pure Yang-Mills action on the Moyal
space exhibits UV/IR mixing \cite{Hayakawa:1999yt,Matusis:2000jf}
which makes its renormalisability quite unlikely unless it is suitably
modified. It can be easily realized that the initial solution proposed
in \cite{Grosse:2004yu} within the real-valued $\varphi^4$-model
cannot be merely extended to gauge theories on Moyal spaces. In the
absence of clear guideline, one reasonable way to follow is to assume
that the Langmann-Szabo duality may appear as a necessary ingredient
in the construction of a renormalisable gauge theory as it has been
the case for the real-valued $\varphi^4$-model. Then, any attempt to
adapt the solution given in \cite{Grosse:2004yu} to gauge theories
would presumably amount to reconcile within a modified action its
invariance under gauge transformations with some covariance under the
Langmann-Szabo duality. More technically, one has to determine whether
or not the naive noncommutative Yang-Mills action can be
supplemented by additional terms that preserve gauge invariance while
making possible the appearance of covariance under the Langmann-Szabo
duality. A convenient way to actually determine all the
above-mentioned additional gauge invariant terms can be achieved by
computing, at least at the one-loop order, the noncommutative
effective gauge theory stemming from a matter field theory coupled to
an external gauge potential in a gauge-invariant way. This is the main
purpose of the present paper.\par

The paper is organised as follows. We start from a renormalisable
scalar (Euclidean) field theory extending to complex-valued fields $\phi$ the
renormalisable noncommutative $\varphi^4$-model with harmonic oscillator
term studied in \cite{Grosse:2004yu,Gurau:2005gd}. This is presented in
section 2 where we also collect the main technical tools. The above
action is minimally coupled to an external gauge potential giving rise
to a gauge-invariant action $S(\phi,A)$. The analysis is based
consistently on the usual algebraic definition of noncommutative
connections for which the modules of the Moyal algebra plays the role
of the set of sections of vector bundles of the ordinary geometry,
while the noncommutative analogue of gauge transformations are
naturally associated with automorphism of (hermitian) modules.
This is presented in detail in the second part of section 2. From
$S(\phi,A)$, we compute the one-loop effective action $\Gamma(A)$
obtained as usual by formally integrating out the scalar field. The
corresponding calculation of the various contributions relevant to the
effective action is presented in section 3. All the computations
are performed within the $x$-space formalism. The resulting action is
further analysed and discussed in section 4. The implications of
the non-vanishing of the one-point (tadpole) contribution are
outlined. This non-vanishing triggers automatically the occurrence of
gauge invariant terms supplementing the noncommutative version of the
pure Yang-Mills term in the effective action. This suggests a possible
expression of a candidate for a renormalisable action for a gauge
theory defined on Moyal spaces in which these additional terms would
be the gauge theory counterpart of the the harmonic term ensuring the
renormalisability of the $\varphi^4$-theory.

\section{External gauge potentials 
coupled to scalar models.}

\subsection{The 4-dimensional complex scalar model.}

We first collect the mathematical tools entering the definition of
the Moyal algebra that will be relevant for the ensuing analysis. A
more mathematical presentation can be found in
\cite{Gracia-Bondia:1987kw,Varilly:1988jk}. In the following, the
``$\star$'' symbol denotes the associative Moyal-Groenenwald product. It
can be first defined on ${\cal{S}}({\mathbb{R}}^4)$ (denoted in short
by ${\cal{S}}$ in the following), the space of complex-valued Schwartz
functions on ${\mathbb{R}}^4$ with fast decay at infinity, by
\begin{align}
(f\star h)(x)={{1}\over{(2\pi)^4}}\int d^4y\,d^4k\ 
f(x+{{1}\over{2}}\Theta.k)\,h(x+y)e^{i k.y},\quad 
\forall f,h\in{\cal{S}},\label{eq:moyal}
\end{align}
such that $(f\star h)\in {\cal{S}}$, where
$\Theta.k\equiv\Theta_{\mu\nu}k^\nu$. Moreover, $\Theta_{\mu\nu}$ is an
invertible constant skew-symmetric matrix which in 4D can be chosen as
$\Theta=\theta{{\Sigma}}$ with
\begin{align}
{{\Sigma}}=\begin{pmatrix} J &0 \\ 0& J \end{pmatrix} , \label{eq:theta}
\end{align}
where\footnote{The above choice for $\Theta_{\mu\nu}$ simplifies
  noticeably the calculation of the effective action. Although this
  choice breaks apparently the $SO(4)$ ``Lorentz'' invariance, it
  turns out that the calculation can be actually performed in a
  Lorentz-covariant way.}  the $2\times2$ matrix $J$ is given by
$J =\begin{pmatrix} 0&-1 \\ 1& 0 \end{pmatrix}$ and the parameter
$\theta$ has mass dimension $-2$. Let ${\cal{S}}^\prime$ denotes the
space of tempered distributions. 
Then, the $\star$-product is further extended to ${\cal{S}}^\prime
\times{\cal{S}}$ upon using duality of linear spaces: 
$\langle T\star f,h \rangle = \langle T,f\star
h\rangle$, $\forall T\in{\cal{S}}^\prime$, $\forall f,h\in{\cal{S}}$,
In a similar way, \eqref{eq:moyal} can be extended
to ${\cal{S}} \times {\cal{S}}^\prime$.
Owing to the smoothening
properties of \eqref{eq:moyal} together with
\begin{align}
\int d^4x\ (f\star h)(x)=\int d^4x\ f(x).h(x), \label{eq:tracial}
\end{align}
where the symbol ``.'' denotes the (commutative) usual pointwise
product, one can show that $T\star f$ and $f \star T$ are smooth functions 
\cite{Gracia-Bondia:1987kw, Varilly:1988jk}.
Now, let ${\cal{L}}$ (resp.\
${\cal{R}}$) denote the subspace of ${\cal{S}}^\prime$ whose
multiplication from right (resp.\ left) by any Schwartz functions is a
subspace of ${\cal{S}}$, namely
\begin{align}
{\cal{L}}=\big\{ T\in{\cal{S}}^\prime~:~~ T\star f\in{\cal{S}}, ~
\forall f\in {\cal{S}}\big\},\quad 
{\cal{R}}=\big\{ T\in{\cal{S}}^\prime ~:~~ 
f\star T\in{\cal{S}}, ~\forall f\in {\cal{S}}\big\}. \label{eq:left-right}
\end{align}
The Moyal algebra, hereafter denoted by ${\cal{M}}$, is then defined as
\begin{align}
{\cal{M}}={\cal{L}}\cap{\cal{R}}.\label{eq:Moyal-alg}
\end{align}
The Moyal algebra is a unital algebra which involves,
in particular, the ``coordinate'' functions $x_\mu$ satisfying
$[x_\mu,x_\nu]_\star =i\Theta_{\mu\nu}$, where this last relation is
well defined on ${\cal{M}}$ ($[a,b]_\star\equiv a\star
b-b\star a$). Other relevant properties of the $\star$-product
that hold on ${\cal{M}}$ are 
\begin{subequations}
\begin{align}
\partial_\mu(f\star h)=\partial_\mu f\star h+f\star\partial_\mu
h,\qquad 
(f\star h)^\dag=h^\dag\star f^\dag,\qquad 
[x_\mu,f]_\star=i\Theta_{\mu\nu}\partial_\nu f, \label{eq:relat1}
\\
x_\mu\star f=(x_\mu.f)+{{i}\over{2}}\Theta_{\mu\nu}\partial_\nu f, 
\qquad 
x_\mu(f\star h)=(x_\mu.f)\star h
-{{i}\over{2}}\Theta_{\mu\nu}f\star\partial_\nu h ,\label{eq:relat2}
\end{align}
\end{subequations}
for any $f,h\in{\cal{M}}$, where in \eqref{eq:relat1} the symbol
$^\dag$ denotes the complex conjugation that permits one to turn
${\cal{M}}$ into an involutive algebra.\par

The action for the (Euclidean) scalar model defined on ${\cal{M}}$
that will be considered in this paper is given by
\begin{equation}
S(\phi)=\int d^4x\big(\partial_\mu\phi^\dag\star\partial_\mu\phi
+\Omega^2(\widetilde{x}_\mu\phi)^\dag\star(\widetilde{x}_\mu\phi)
+m^2\phi^\dag\star\phi\big)(x)+S_{int},\label{eq:actionharm}
\end{equation}
where $\phi$ is a complex scalar field with mass $m$, $S_{int}$
denotes the interaction terms to be discussed below and we have set
${\widetilde{x}}_\mu=2\Theta^{-1}x$. The parameters $\Omega$ and $\lambda$
are dimensionless. At this point, some comments are in order. This
model cannot be viewed as related to some LSZ-type model
\cite{Langmann:2003if,Langmann:2003cg} since in that latter case the
corresponding action would have been of the form
\begin{equation}
S_{LSZ}(\phi)=\int d^4x\big((\partial_\mu\phi
+i\Omega{\widetilde{x}}_\mu\phi)^\dag\star
(\partial_\mu\phi+i\Omega{\widetilde{x}}_\mu\phi)
+m^2\phi^\dag\star\phi\big)(x)+S_{int}.\label{eq:actionlsz}
\end{equation}
It can easily be realised that the quadratic terms in
\eqref{eq:actionlsz} do not coincide with those involved in
\eqref{eq:actionharm}, giving rise therefore to different propagators
for these actions (as well as, anticipating with the discussion of the
next subsection, different minimal coupling prescriptions). Notice
however that both actions are covariant under the Langmann-Szabo
duality \cite{Langmann:2002cc}. It turns out, as it will be shown in a while,
that the operator $\partial_\mu + i\Omega{\widetilde{x}}_\mu$ can actually
be related to a connection $\nabla^\zeta_\mu$ with
$\zeta = -{{\Omega}\over{1+\Omega}}{\widetilde{x}}_\mu$. In
\eqref{eq:actionharm}, the term involving $\Omega$ can be viewed as
the (complex-valued) scalar counterpart of the harmonic oscillator
term first introduced in \cite{Grosse:2004yu} leading to the construction
of a renormalisable noncommutative (real-valued) $\varphi^4_4$-model.

Although our one-loop computation of effective actions will not depend
on the explicit form of the interaction, it is instructive to discuss
it more closely in view of the corresponding consequences on the
renormalisability of the models. The most general interaction can be
written as
\begin{equation}
S_{int}=S_{int}^0+S_{int}^{NO}
=\int \lambda(\phi^\dag\star\phi\star\phi^\dag\star\phi)(x)
+\kappa(\phi^\dag\star\phi^\dag\star\phi\star\phi)(x).\label{eq:scalarint}
\end{equation}
We point out that the only diagrams that can be orientated are those
occurring in the loopwise expansion obtained from $S_{int}^O$ while
$S_{int}^{NO}$ yields diagrams in the loopwise expansion that cannot
be oriented. Recall now that the proof of the renormalisability of the
noncommutative version of the Gross-Neveu model studied in
\cite{Vignes-Tourneret:2006nb} (whose interaction term is the
fermionic counterpart of $S_{int}^O$) relies heavily on the
orientability of the diagrams.  It turns out \cite{Gurau:2005gd} that
\eqref{eq:actionharm} restricted to $S_{int}^O$ is renormalisable for
any value of $\Omega$. Besides, a similar conclusion applies for the
LSZ-type model \eqref{eq:actionlsz} restricted to $S_{int}^O$. The
proof, as sketched in \cite{Gurau:2005gd}, is somehow similar to the
one given in \cite{Vignes-Tourneret:2006nb} for the noncommutative
Gross-Neveu model. At the present time, the actual impact of
interaction terms as given by $S_{int}^{NO}$ on the renormalisability
of the above models is not known.\par

\begin{figure}[!htb]
  \centering
  \includegraphics[scale=1]{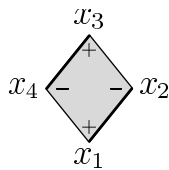}
  \caption[The Vertex]{\footnotesize{Graphical representation for the
      vertex in the $x$-space, obtained from \eqref{eq:interaction}.
      The plus-sign (resp.\ minus-sign) appearing in the rhombus
      corresponds to incoming (resp.\ outgoing) external line
      associated with ${\phi^\dag}$ (resp.\ $\phi$).}}
  \label{fig:vertex}
\end{figure}

The Feynman graphs can be computed from the propagator and interaction
vertex derived from \eqref{eq:actionharm}. In the following, we will
work within the $x$-space formalism \cite{Gurau:2005gd} which proves
particularly convenient as it simplifies the calculations. The scalar
propagator $C(x,y) \equiv \langle\phi(x)\phi^\dag(y)\rangle$ in 
$x$-space obtained by solving
$(\Delta_x+{\widetilde{\Omega}}^2x^2+m^2)C(x,y) =\delta(x - y)$ is
given by
\begin{equation}
C(x,y)={{\Omega^2}\over{\pi^2\theta^2}}\int_0^\infty \!\! 
{{dt}\over{\sinh^2(2{\widetilde{\Omega}}t)}}\exp\Big(
-{{{\widetilde{\Omega}}}\over{4}}\coth({\widetilde{\Omega}}t)(x{-}y)^2
-{{{\widetilde{\Omega}}}\over{4}}\tanh({\widetilde{\Omega}}t)(x{+}y)^2-m^2t\Big),
\label{eq:propag}
\end{equation}
where we have defined ${\widetilde{\Omega}}\equiv
2{{\Omega}\over{\theta}}$. The interaction vertices can be read off
from the RHS of
\begin{subequations}
  \label{eq:interaction}
  \begin{align}
    \int d^4x(\phi^\dag\star\phi\star\phi^\dag\star\phi)(x)
&=\frac{1}{\pi^4\theta^4}\int 
    \prod_{i=1}^4d^4x_i\,\phi^\dag(x_1)\phi(x_2)\phi^\dag(x_3)\phi(x_4)
\label{eq:InteractionDvpee}
\\
 &\times\delta(x_1-x_2+x_3-x_4)e^{-i\sum_{i<j}(-1)^{i+j+1}x_i\wedge
   x_j }. \nonumber
    \intertext{We will denote the vertex kernel as}
    V(x_1,x_2,x_3,x_4)&=\delta(x_1-x_2+x_3-x_4)
e^{-i\sum_{i<j}(-1)^{i+j+1}x_i\wedge x_j 
    } \label{eq:InteractionKernel}
  \end{align}
\end{subequations}
in which $x\wedge y\equiv 2x_\mu\Theta^{-1}_{\mu\nu}y_\nu$. The
generic graphical representation of the vertex is depicted on the
figure~\ref{fig:vertex}. The non-locality of the interaction is
conveniently represented by the rhombus appearing on
fig.~\ref{fig:vertex} whose vertices correspond to the $x_i$'s
occurring in \eqref{eq:interaction}. It is useful to represent the
alternate signs in the delta function of \eqref{eq:interaction} by
plus- and minus-signs, as depicted on the figure. By convention, a
plus-sign (resp.\ minus-sign) corresponds to an incoming field
$\phi^\dag$ (resp.\ outgoing field $\phi$). This permits one to define
an orientation on the diagrams obtained from the loop expansion.

\subsection{Gauge connexions on Moyal Space.}

It is necessary to define clearly the mathematical status
\cite{CONNES,CM} (see also \cite{Dubois-Violette:1989vq,Gayral:2006wu})
of the various objects that will be involved in the minimal coupling
prescription. Recall that ${\cal{M}}$ is a unital involutive algebra.
Let ${\cal{H}}$ be a right ${\cal{M}}$-module with hermitian structure
$h$, that is, a sesquilinear map $h:{\cal{H}}\times{\cal{H}} \to
{\cal{M}}$ such that $h(m_1\star f_1,m_2\star f_2) = f_1^\dag\star
h(m_1,m_2)\star f_2$, for any $f_1,f_2 \in {\cal{M}}$, and $m_1,m_2
\in {\cal{H}}$. The algebra ${\cal{M}}$ is assumed to be endowed with
a differential calculus based on the derivations $\partial_\mu$. The
usual concept of connections defined on vector bundles in ordinary
geometry can be consistently generalised in noncommutative geometry to
connections on projective modules (over an associative algebra).
Namely, a connection can be defined (algebraically) by a linear map
$\nabla_\mu: {\cal{H}} \to {\cal{H}}$ verifying the Leibnitz rule:
\begin{align}
\nabla_\mu(m\star f)=\nabla_\mu(m)\star
f+m\star\partial_\mu f,\ \forall m\in{\cal{H}},\quad 
\forall f\in{\cal{M}}\label{eq:nabla}
\end{align} 
and preserving the hermitian structure, that is
\begin{align}
\partial_\mu h(m_1,m_2)=h(\nabla_\mu m_1,m_2)+h(m_1,\nabla_\mu
m_2),\quad \forall m_1,m_2\in{\mathcal{H}}. \label{eq:hermite}
\end{align}
When ${\cal{H}} = {\cal{M}}$, that
we assume from now on, it follows from \eqref{eq:nabla} that the connection is
entirely determined by its action $\nabla_\mu(\mathbb{I})$ on the unit
$\mathbb{I} \in \mathcal{M}$, denoted by
\begin{align}
\nabla_\mu^A(\mathbb{I})\equiv-iA_\mu,\label{eq:connection-def}
\end{align}
since one has obviously $\nabla^A_\mu(\mathbb{I}\star
f) = \nabla^A_\mu(\mathbb{I})\star f+\partial_\mu
f \equiv \partial_\mu f-iA_\mu\star f$.  This therefore represents
the gauge potential $A_\mu$ in ${\cal{M}}$. Observe that for
${\cal{H}} = {\cal{M}}$, a hermitian structure is provided by
$h(f_1,f_2) = f_1^\dag\star f_2$, ensuring that the above connections
are hermitian.\par

Gauge transformations hereafter denoted by $\gamma$ are determined by
automorphisms of the module ${\cal{M}}$ (keeping in mind that
${\cal{M}}={\cal{H}}$ is considered as a hermitian module over itself)
preserving the hermitian structure $h$, $\gamma \in Aut_h({\cal {M}})$.
One has{\footnote{When ${\cal{H}} \ne {\cal{M}}$, recall that
    $\gamma$, as a morphism of module, satisfies $\gamma(m\star f) =
    \gamma(m)\star f$, $\forall m \in {\cal{H}}$, $\forall f \in
    {\cal{M}}$.}}
\begin{align}
\gamma(f) &=\gamma(\mathbb{I}\star
f)=\gamma(\mathbb{I})\star f\ ,\quad \forall f\in{\mathcal{M}},
\nonumber
\\
h\big(\gamma(f_1),\gamma(f_2)\big) &=
h(f_1,f_2) \quad \forall f_1,f_2 \in \mathcal{M} \qquad \Rightarrow 
\quad \gamma(\mathbb{I})^\dag \star \gamma(\mathbb{I}) = \mathbb{I},
\label{eq:unitary}
\end{align}
so that gauge transformations are entirely determined by
$\gamma(\mathbb{I})\in {\cal{U}}(\cal{M})$, where ${\cal{U}}(\cal{M})$
is the group of unitary elements of $\cal{M}$. From now on, we set
$\gamma(\mathbb{I}) \equiv g$. Then, according to
\eqref{eq:unitary}, the action of the gauge group on any matter field
$\phi \in {\cal{M}}$ can be defined by
\begin{equation}
\phi^g=g\star\phi\label{eq:fundam}
\end{equation}
for any $g \in {\cal{U}}(\cal{M})$, which may be viewed, in more
physical words, as the noncommutative analogue of the transformation
of the matter fields under the ``fundamental representation of the
gauge group''. Note that one has $g^\dag\star g = g\star g^\dag =
\mathbb{I}$.\par

The action of ${\cal{U}}({\cal {M}})$ on the connection $\nabla^A_\mu$ is
given by
\begin{align}
(\nabla^A_\mu)^\gamma(\phi)
=\gamma(\nabla^A_\mu(\gamma^{-1}\phi)),\quad 
\forall \phi\in{\cal{M}}. \label{eq:transjauge}
\end{align}
By further using $\gamma(\phi) = \gamma(\mathbb{I} \star \phi) = g
\star \phi$ together with the expression of the covariant derivative
\begin{align}
\nabla_\mu^A(\phi)=\partial_\mu\phi-iA_\mu\star\phi \label{eq:covder}
\end{align}
and the fact that $(\nabla_\mu^A)^g \equiv\partial_\mu-iA^g_\mu$, one
obtains the following gauge transformation for the gauge potential
$A_\mu$
\begin{align}
A_\mu^g=g\star A_\mu\star g^\dag+ig\star\partial_\mu g^\dag .
\label{eq:gaugetransf}
\end{align}
In the present noncommutative (algebraic) framework, the space of
gauge potentials $A_\mu\in\cal{M}$ is a linear space (this comes
basically from the fact that ${\cal{M}}$, as a module, is a linear
space). Note that any one-form can be used to define a connection so
that if some $A_\mu$ defines a connection, then $\lambda A_\mu$,
$\forall\lambda \in {\mathbb{R}}$, defines another connection. There
is a subtlety here that must be pointed out. The gauge transformations
\emph{do not preserve the structure of linear space} of gauge
potentials since 
\begin{align}
(\lambda A_\mu)^g-\lambda(A_\mu^g)= i(1-\lambda)g\star\partial_\mu
g^{\dag}. 
\end{align}
This is easily obtained by comparing how the gauge transformations as
given by \eqref{eq:transjauge} operate on $\nabla_\mu^{\lambda A}$ and
$\nabla_\mu^A$ according to \eqref{eq:covder} and expresses the fact
that multiplication of a gauge potential by a scalar and 
gauge transformation are two noncommuting operations. The same
discussion applies to the sum of two gauge potentials $A_1+A_2$. \par

It is useful to exhibit a special reference connection that will play
a salient role in the following. It turns out that
\begin{align}
\xi_\mu\equiv-{{1}\over{2}}\widetilde{x}_\mu \label{eq:invarcon} 
\end{align}
defines a connection invariant under gauge transformations. Note that
the occurrence of gauge-invariant connections is not new in
noncommutative geometry and has been already mentioned in earlier
studies focused in particular on matrix-valued field theories
\cite{Dubois-Violette:1989vq,Dubois-Violette:1998,Masson:1999,Masson:2005}.
Indeed, according to \eqref{eq:covder}, the connection
$\nabla_\mu^\xi$ associated to $\xi_\mu$ verifies
\begin{align}
\nabla^\xi_\mu\phi=\partial_\mu\phi-i\xi_\mu\star\phi
=-i\phi\star\xi_\mu,\label{eq:inconnu}
\end{align}
where the second equality stems from the following relation
\begin{align}
\partial_\mu\phi=[i\xi_\mu,\phi]_\star \label{eq:innerder}
\end{align}
which simply expresses the fact that the derivative $\partial_\mu$ in
$\cal{M}$ is an inner derivative. Then, as $\nabla_\mu^\xi$ given by
right multiplication commutes with the gauge transformation
\eqref{eq:transjauge} given by left multiplication, it is easy to
realise that
\begin{align}
(\nabla_\mu^\xi)^g(\phi)= g\star(\nabla_\mu^\xi(g^\dag\star\phi))
=-i\phi\star\xi_\mu=\nabla_\mu^\xi\phi.
\end{align}
The second equality stems from \eqref{eq:inconnu}, which shows that
the connection $\nabla_\mu^\xi$ is invariant under the gauge
transformations, from which follows that
\begin{align}
  \xi_\mu^g=\xi_\mu,\label{eq:invar-xi}
\end{align}
as it could have been checked directly by combining the actual
expression for $\xi_\mu$ with \eqref{eq:gaugetransf} and
\eqref{eq:innerder}. In the present Moyal framework, the existence of
the above invariant connection seems to be an unavoidable consequence
of the existence of inner derivations{\footnote{One of us (J.C.W) is
    grateful to M. Dubois-Violette for an enlightening discussion on
    this point.}} as defined by \eqref{eq:innerder} (it turns out that
all derivations on the Moyal algebra are inner derivations).\par

Let us introduce now
\begin{align}
\nabla^A_\mu-\nabla^\xi_\mu=-i(A_\mu-\xi_\mu)\equiv-i\mathcal{A}_\mu
\label{eq:arondvrai}
\end{align}
which, as the difference of two connections, defines obviously a
tensorial form ${\cal{A}}_\mu$ whose gauge transformations are given
by
\begin{align}
\mathcal{A}_\mu^g=g\star\mathcal{A}_\mu\star g^\dag.\label{eq:arondfaux}
\end{align}
This tensorial form has been sometimes called in the String Theory
literature the covariant coordinates (see e.g.\ \cite{Douglas:2001ba}
and references therein). Given a connection $\nabla_\mu^A$ (or
equivalently a gauge potential $A_\mu$), the corresponding curvature
is given by
\begin{align}
F_{\mu\nu}^A=i[\nabla^A_\mu,\nabla^A_\nu]_\star=
\partial_\mu A_\nu-\partial_\nu A_\mu-i[A_\mu,A_\nu]_\star ,
\label{eq:curvature}
\end{align}
with gauge transformations taking the usual form
\begin{align}
(F_{\mu\nu}^A)^g=g\star F_{\mu\nu}^A\star g^\dag.
\end{align}
By further combining \eqref{eq:curvature} with \eqref{eq:innerder} and
\eqref{eq:arondvrai}, the curvature can be reexpressed as
\begin{align}
F_{\mu\nu}^A=\Theta_{\mu\nu}^{-1}
-i[\mathcal{A}_\mu,\mathcal{A}_\nu]_\star.\label{eq:f-arond}
\end{align}
Note that the invariant connection defined by $\xi_\mu$ is a constant
curvature connection since $F_{\mu\nu}^\xi =
\Theta_{\mu\nu}^{-1}$.\par

Another type of transformations given by $\phi^U = U\star \phi\star
U^\dag$, which may be viewed as the noncommutative analogue of
transformations of matter fields in the adjoint representation, has
been also considered in the literature. These transformations will be
more closely analysed in the next subsection.\par

\subsection{The minimal coupling prescription.}

Let us assume that the action of the gauge group on the matter fields
$\phi$ is given by \eqref{eq:fundam}. Then, owing to the special role
played by the coordinate functions $x_\mu$ through the invariant
``gauge potential'' \eqref{eq:invarcon} involved in $\nabla^\xi_\mu$ and
the expression for the inner derivatives \eqref{eq:innerder}, it
follows that a natural choice for the minimal coupling of the action
\eqref{eq:actionharm} to an external gauge field $A_\mu$ is obtained
by performing the usual substitution
\begin{align}
\partial_\mu\to\nabla_\mu^A
\end{align}
on the action \eqref{eq:actionharm} {\it{provided this latter is
    reexpressed in terms of}} $\partial_\mu$ {\it{and}}
$\nabla_\mu^\xi$, using in particular the following identity:
\begin{align}
\widetilde{x}_\mu\phi=\widetilde{x}_\mu\star\phi-i\partial_\mu\phi 
=-i(\partial_\mu\phi-2i\xi_\mu\star\phi)
= -2i\nabla_\mu^\xi\phi+i\partial_\mu\phi. \label{eq:identitycouplage}
\end{align}
By using \eqref{eq:identitycouplage}, one easily infers that the
minimal coupling prescription can be conveniently written as
\begin{align}
\partial_\mu\phi&\mapsto \nabla_\mu^A\phi=
\partial_\mu\phi-iA_\mu\star\phi, \label{eq:coup1}\\
\widetilde{x}_\mu\phi &\mapsto 
-2i\nabla_\mu^\xi\phi+i\nabla_\mu^A\phi=\widetilde{x}_\mu\phi+
A_\mu\star\phi.\label{eq:coup2}
\end{align}
Note that gauge invariance of the resulting action functional is
obviously obtained thanks to the relation
$(\nabla^{A,\xi}_\mu(\phi))^g = g\star(\nabla^{A,\xi}_\mu(\phi))$.\par

By applying the above minimal coupling prescription to
\eqref{eq:actionharm}, we obtain the following gauge-invariant action
\begin{align}
S(\phi,A) =& S(\phi)+ \int d^4x \ \big((1+\Omega^2)\phi^\dag\star
(\widetilde{x}_\mu A_\mu)\star\phi\nonumber\\ 
&-(1-\Omega^2)\phi^\dag\star A_\mu \star\phi\star
\widetilde{x}_\mu+
(1+\Omega^2)\phi^\dag\star A_\mu\star
A_\mu\star \phi\big)(x),\label{eq:harmcoupled}
\end{align}
where $S(\phi)$ is given by \eqref{eq:actionharm} with 
$S_{int}$ restricted to its gauge-invariant part
$S_{int}^O$, see \eqref{eq:scalarint}. At this level, it is
instructive to interpret the action \eqref{eq:actionlsz} in the light
of the algebraic framework that has been developed above. As already
mentioned in subsection 2.1, the operator
$\partial_\mu+i\Omega{\widetilde{x}}_\mu$ is actually related to a
connection $\nabla^{\zeta}_\mu$ with
\begin{align}
\zeta_\mu=\frac{2\Omega}{1+\Omega}\xi_\mu ,\label{eq:zeta}
\end{align}
since the following relation
\begin{align}
(\partial_\mu+i\Omega{\widetilde{x}}_\mu)\phi 
=(1+\Omega)\Big(\partial_\mu\phi
-i{{2\Omega}\over{1+\Omega}}\xi_\mu\star\phi\Big) 
=(1+\Omega)\nabla^{\zeta}_\mu(\phi)
\end{align}
holds in view of \eqref{eq:covder}. The action \eqref{eq:actionlsz}
can then be rewritten as
\begin{align}
S_{LSZ}(\phi)=\int d^4x\big((1+\Omega)^2
(\nabla^{\zeta}_\mu(\phi))^\dag\star\nabla^{\zeta}_\mu(\phi)
+m^2\phi^\dag\star\phi\big)(x)+S_{int},
\end{align}
where $\zeta$ is given by \eqref{eq:zeta} which, for $\Omega \ne 0$,
makes explicit the invariance of the action under the gauge
transformations $\phi^g = g\star\phi$ for any
$g\in{\cal{U}}({\cal{M}})$. Notice that a similar comment applies to
the noncommutative version of the (two-dimensional) Gross-Neveu model
considered recently in \cite{Vignes-Tourneret:2006nb}. It can be
easily realised that the corresponding action quoted in
\cite{Vignes-Tourneret:2006nb} can be cast into the form
\begin{align}
S_{GN}=\int d^2x\big(-i(1+\Omega){\bar{\psi}}\gamma^\mu
\nabla_\mu^\zeta\psi+m{\bar{\psi}}\psi \big)(x)+...,\label{eq:GN}
\end{align}
where the ellipses denote interaction terms, $\psi$ is a spinor and
the antihermitian $\gamma$ matrices satisfy $\{\gamma_\mu,\gamma_\nu\}
= -2\delta_{\mu\nu}$. In physical words, it should be clear that these
two latter actions can be interpreted as matter actions already
coupled to an external (background) gauge potential $\zeta_\mu$ (while
the action \eqref{eq:actionharm} does not obviously support this
interpretation).\par

As announced in the last subsection, another type of transformations
given by
\begin{align}
\phi^U=U\star \phi\star U^\dag\equiv\alpha(\phi),\label{eq:adjointde}
\end{align}
for any $U\in{\cal{U}}({\cal{M}})$, has been also considered in the
literature. It is instructive to confront the actual mathematical
status of these transformations to the algebraic framework developed
in subsection 2.2. In fact, it should be clear that
\eqref{eq:adjointde} defines an automorphism $\alpha$ of algebra,
\begin{align}
  \alpha(\phi_1\star\phi_2)=\alpha(\phi_1)\star\alpha(\phi_2),
  \label{eq:automalg}
\end{align} {\it{but not}} an automorphism of the module (which would
satisfy $\alpha(\phi_1\star\phi_2) = \alpha(\phi_1)\star \phi_2$)
except when $U$ is in the centre of ${\cal{M}}$ (which in the present
case is equal to ${\mathbb{C}}$).  Actually, the noncommutative
analogue of the adjoint representation of the gauge group is
constructed with the help of the real structure $J$ (see e.g.\ 
\cite{CM}). This requires to replace the algebra
$\mathcal{M}$ by $\mathcal{M} \otimes \mathcal{M}^o$, where
$\mathcal{M}^o$ is the opposite algebra.  The only minimal coupling
prescription which is compatible with modules over the algebra
$\mathcal{M}$ is given by \eqref{eq:coup1}, \eqref{eq:coup2}.\par

Nonetheless, in order to prepare the discussion of section 4, we
simply quote the action 
\begin{align}
S_{adj}(\phi,A)=& S(\phi)+ \int d^4x \ \big((1+\Omega^2)(\phi^\dag\star
(\widetilde{x}_\mu A_\mu)\star\phi+ \phi\star
(\widetilde{x}_\mu A_\mu)\star\phi^\dag)\nonumber\\
&-(1-\Omega^2)(\phi^\dag\star A_\mu \star\phi\star
\widetilde{x}_\mu
+ \phi\star A_\mu \star\phi^\dag\star
\widetilde{x}_\mu)-2(1-\Omega^2)\phi^\dag\star A_\mu\star
\phi\star A_\mu \nonumber\\
 &+(1+\Omega^2)(\phi^\dag\star A_\mu\star
A_\mu\star \phi
+ \phi\star A_\mu\star A_\mu\star
\phi^\dag)\big)(x),\label{eq:adjcoupled}
\end{align}
which is invariant under the adjoint gauge transformation
(\ref{eq:adjointde}). This is obtained from \eqref{eq:actionharm} by the
substitution
\begin{align}
\partial_\mu\phi \mapsto
\partial_\mu\phi-i[A_\mu ,\phi ]_\star\,,\qquad 
\widetilde{x}_\mu\phi \mapsto \widetilde{x}_\mu\phi+
\{A_\mu ,\phi\}_\star\,.
\end{align}

\par

\section{The one-loop effective action.}

In this section we will calculate the one-loop effective action
starting from the action $S(\phi,A)$ \eqref{eq:harmcoupled}. Recall
that the effective action is formally obtained from
\begin{align}
e^{-\Gamma(A)}\equiv \int D\phi D\phi^\dag e^{-S(\phi,A)}
=\int D\phi D\phi^\dag e^{-S(\phi)} e^{-S_{int}(\phi,A)}, \label{eq:defact}
\end{align}
where $S(\phi)$ is given by \eqref{eq:actionharm} and
$S_{int}(\phi,A)$ can be read off from \eqref{eq:harmcoupled} and
\eqref{eq:actionharm}. At the one-loop order, \eqref{eq:defact}
reduces to
\begin{align}
e^{-\Gamma_{1loop}(A)}=\int D\phi D\phi^\dag e^{-S_{free}(\phi)} 
e^{-S_{int}(\phi,A)},
\end{align}
where $S_{free}(\phi)$ is simply the quadratic part of
\eqref{eq:actionharm}. The corresponding diagrams are depicted on the
figures \ref{fig:1-point}-\ref{fig:4-point}.\par

\begin{figure}[!htb]
  \centering
  \includegraphics[scale=1]{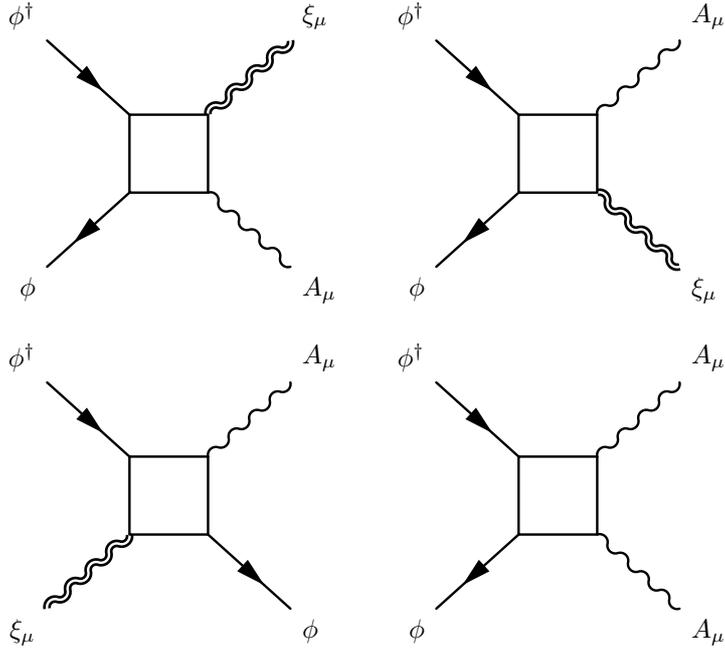}
  \caption[The vertices]{\footnotesize{Graphical representation for
      the vertices carrying the external gauge potential $A_\mu$
      involved in the action \eqref{eq:harmcoupled}. The overall
      factor affecting the two uppermost vertices is $(1 + \Omega^2)$.
      From left to right, the overall factors affecting the lower
      vertices are respectively equal to $- 2(1 - \Omega^2)$ and $- (1
      + \Omega^2)$.}}
  \label{fig:vertices}
\end{figure}

The additional vertices involving $A_\mu$ and/or $\xi_\mu$ and
generated by the minimal coupling can be obtained by combining
\eqref{eq:interaction} with \eqref{eq:harmcoupled} and the generic
relation
\begin{align}
\int d^4x(f_1\star f_2\star f_3\star f_4)(x)=\frac{1}{\pi^4\theta^4}\int 
\prod_{i=1}^4d^4x_i\, f_1(x_1) f_2(x_2) f_3(x_3) f_4(x_4) \nonumber\\
\times\delta(x_1-x_2+x_3-x_4)e^{-i\sum_{i<j}(-1)^{i+j+1}x_i\wedge x_j }.
\end{align}
These vertices are depicted on the figure \ref{fig:vertices}. Note
that additional overall factors must be taken into account. These are
indicated on the figure \ref{fig:vertices}.

\subsection{The tadpole for the scalar model.}

\begin{figure}[!htb]
  \centering
  \includegraphics[scale=1]{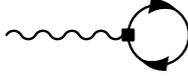}
  \caption[1-point]{\footnotesize{The non vanishing tadpole diagram.
      To simplify the figure, we do not explicitly draw all the
      diagrams that would be obtained from the vertices given on the
      figure 2 but indicate only the overall topology of the
      corresponding diagrams. Notice that the background lines are not
      explicitly depicted.}}
  \label{fig:1-point}
\end{figure}
Using the expression for the vertices and the minimal coupling, the
amplitude corresponding to the tadpole on figure \ref{fig:1-point} is
\begin{equation}
\mathcal{T}_1= \frac{1}{\pi^4\theta^4}\int d^4x\ d^4u\ d^4z\ A_\mu(u)\  
e^{-i(u-x)\wedge z}\ C(x+z,x)\  
((1-\Omega^2)(2\widetilde{x}_\mu+\widetilde{z}_\mu)-2\widetilde{u}_\mu). 
\label{eq:tadpole1}
\end{equation}
Combining this with the explicit expression for the propagator
\eqref{eq:propag}, \eqref{eq:tadpole1} can be expressed as
\begin{align}
\mathcal{T}_1=& \frac{\Omega^2}{4\pi^6\theta^6}\int d^4x\ d^4u\ d^4z 
\int_0^\infty  \frac{dt\ e^{-tm^2}}{ \sinh^2(\widetilde{\Omega}t)
\cosh^2(\widetilde{\Omega}t)}\  A_\mu(u)\  e^{-i(u-x)\wedge z}\nonumber\\
&\times e^{-\frac{\widetilde{\Omega}}{4} (\coth(\widetilde{\Omega}t)z^2
+\tanh(\widetilde{\Omega}t)(2x+z)^2}
((1-\Omega^2)(2\widetilde{x}_\mu+ \widetilde{z}_\mu)-2\widetilde{u}_\mu).
\label{eq:tadpole2}
\end{align}
At this point, we find convenient to introduce the following
8-dimensional vectors $X$, $J$ and the $8\times 8$ matrix $K$ defined
by
\begin{equation}
X=\begin{pmatrix} x\\ z \end{pmatrix}, \quad 
K=\begin{pmatrix} 4\tanh(\widetilde{\Omega}t) \mathbb{I} & 
2\tanh(\widetilde{\Omega}t)\mathbb{I} -2i\Theta^{-1} \\  
2\tanh(\widetilde{\Omega}t)\mathbb{I} +2i\Theta^{-1} &  
(\tanh(\widetilde{\Omega}t)+ \coth(\widetilde{\Omega}t))\mathbb{I}
\end{pmatrix} ,\quad
\ J=\begin{pmatrix} 0\\ i\tilde{u} \end{pmatrix}. \label{eq:tadpole2bis}
\end{equation}
This permits one to reexpress \eqref{eq:tadpole2} in a form such that
some Gaussian integrals can be easily performed. Note that this latter
procedure can be adapted to the calculation of the higher order Green
functions (see subsection 3.2). The combination of \eqref{eq:tadpole2bis}
with \eqref{eq:tadpole2} then yields
\begin{align}
\mathcal{T}_1=& \frac{\Omega^2}{4\pi^6\theta^6}
\int d^4x\ d^4u\ d^4z \int_0^\infty  
\frac{dt\ e^{-tm^2}}{
  \sinh^2(\widetilde{\Omega}t)\cosh^2(\widetilde{\Omega}t)}\  
A_\mu(u) \nonumber\\
&\times e^{-\frac{1}{2}X.K.X+J.X} 
((1-\Omega^2)(2\widetilde{x}_\mu+ \widetilde{z}_\mu)-2\widetilde{u}_\mu).
\end{align}
By performing the Gaussian integrals on $X$, we find
\begin{equation}
\mathcal{T}_1=-\frac{\Omega^4}{\pi^2\theta^2(1+\Omega^2)^3}
\int d^4u \int_0^\infty  \frac{dt\ e^{-tm^2}}{ 
\sinh^2(\widetilde{\Omega}t)\cosh^2(\widetilde{\Omega}t)}\ 
A_\mu(u)\widetilde{u}_\mu\ e^{-\frac{2\Omega}{\theta(1+\Omega^2)}
\tanh(\widetilde{\Omega}t)u^2}. \label{eq:tadpole3}
\end{equation}
Then, inspection of the behaviour of \eqref{eq:tadpole3} for $t \to
0$ shows that this latter expression has a quadratic as well as a
logarithmic UV divergence. Indeed, by performing a Taylor expansion of
\eqref{eq:tadpole3}, one obtains
\begin{align}
\mathcal{T}_1 =& -\frac{\Omega^2}{4\pi^2(1+\Omega^2)^3}
\left( \int d^4u\ \widetilde{u}_\mu A_\mu(u)\right)\ 
\frac{1}{\epsilon}\ -\frac{m^2\Omega^2}{4\pi^2(1+\Omega^2)^3}
\left( \int d^4u\ \widetilde{u}_\mu A_\mu(u)\right)\ \ln(\epsilon) 
\nonumber\\
 & -\frac{\Omega^4}{\pi^2\theta^2(1+\Omega^2)^4}
\left( \int d^4u\ u^2\widetilde{u}_\mu A_\mu(u)\right)\ 
 \ln(\epsilon)\ + \dots,\label{eq:tasdepaul}
\end{align}
where $\epsilon \to 0$ is a cut-off and the ellipses denote finite
contributions. The fact that the tadpole is (a priori) non-vanishing
is a rather unusual feature for a Yang-Mills type theory. This will be
discussed more closely in section 4.\par

\subsection{The multi-point contributions.}

The 2, 3 and 4-point functions can be computed in a way similar to
the one used for the tadpole. The algebraic manipulations are standard
but cumbersome so that we only give below the final expressions for
the various contributions. \par
\begin{figure}[!htb]
  \centering
  \includegraphics[scale=1]{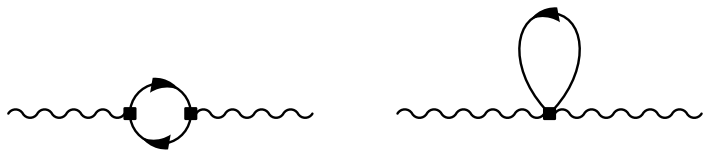}
  \caption[Two-point]{\footnotesize{Relevant one-loop diagrams
      contributing to the two-point function. To simplify the figure,
      we do not explicitly draw all the diagrams that would be
      obtained from the vertices given in figure 2 but indicate
      only the overall topology of the corresponding diagrams. Notice
      that the background lines are not explicitly depicted. The
      leftmost (resp.\ rightmost) diagram corresponds to the
      contribution $\mathcal{T}_2'$ (resp.\ $\mathcal{T}_2''$).}}
  \label{fig:2-point}
\end{figure}
Let us start with the two-point function. The regularisation of the
diverging amplitudes is performed in a way that preserves gauge
invariance of the most diverging terms (which in four dimensions are
UV quadratically diverging) so that the cut-off $\epsilon$ to be put
on the various integrals over the Schwinger parameters, says
$\int_{\epsilon}^\infty dt$, must be suitably chosen. In the present
case, we find that this can be achieved with $\int_{\epsilon}^\infty
dt$ for $\mathcal{T}_2''$ while for $\mathcal{T}_2'$ the
regularisation must be performed with $\int_{\epsilon/4}^\infty$.
Such an adaptation by hand of the scheme is not surprising. The
one-loop effective action can be expressed in terms of heat
kernels \cite{Gayral:2004cs},
\begin{align}
\Gamma_{1loop}(\phi,A)&=-\frac{1}{2} \int_0^\infty \frac{dt}{t} \,
\mathrm{Tr}\big( e^{-t  H(\phi,A)} - e^{-t H(0,0)}\big)
\\
&= -\frac{1}{2} \lim_{s \to 0} \Gamma(s)\,\mathrm{Tr}\big(
H^{-s}(\phi,A)-H^{-s}(0,0)\big),
\nonumber
\end{align}
where $H(\phi,A)=\frac{\delta^2 S(\phi,A)}{\delta \phi \,\delta
  \phi^\dag}$. Expanding \cite{Connes:2006qj}
\begin{align}
H^{-s}(\phi,A) =\big(1+a_1(\phi,A) s + a_2(\phi,A) s^2 + \dots\Big)  
H^{-s}(0,0),
\end{align}
we obtain
\[
\Gamma_{1loop}(\phi,A)
= -\frac{1}{2} \lim_{s \to 0} \mathrm{Tr}\Big(\big(
\Gamma(s{+}1)a_1(\phi,A) + s \Gamma(s{+}1) a_2(\phi,A) 
+\dots \big) H^{-s}(0,0)\Big).
\]
With $\Gamma(s+1)=1-s\gamma+\dots$ we have
\begin{align}
\Gamma_{1loop}(\phi,A)
&= -\frac{1}{2} \lim_{s \to 0} \mathrm{Tr}\big(
a_1(\phi,A) H^{-s}(0,0)\big) 
\nonumber
\\
&-\frac{1}{2} \mathrm {Res}_{s=0} \,\mathrm{Tr} 
\Big( \big(a_2(\phi,A) -\gamma a_1(\phi,A)\big)  H^{-s}(0,0)\Big).
\end{align}
The second line is the Wodzicki residue \cite{Wodzicki:1984}, which is
a trace and corresponds to the logarithmically divergent part of the
one-loop effective action. But there is also the quadratically
divergent part $ -\frac{1}{2} \lim_{s \to 0} \mathrm{Tr}\big( a_1
H^{-s}(0,0)\big)$ in the action which cannot be gauge-invariant.  In
field-theoretical language, gauge invariance is broken by the
naive $\epsilon$-regularisation of the Schwinger integrals and
must be restored by adjusting the regularisation scheme using methods
from algebraic renormalisation \cite{Piguet:1995er}.  In would be
interesting to check that algebraic renormalisation methods leads
indeed to the replacement $\epsilon \mapsto \frac{\epsilon}{4}$ in
$\mathcal{T}_2'$. Note that the logarithmically divergent part is
insensitive to a finite scaling of the cut-off.\par

After some tedious calculations, we find the following final
expressions for the diagrams on figure \ref{fig:2-point} are
\begin{subequations}
\begin{align}
\mathcal{T}_2' &= \frac{(1{-}\Omega^2)^2}{16\pi^2(1{+}\Omega^2)^3}
\left(\int \! d^4u\ A_\mu(u)A_\mu(u)\right) \frac{1}{\epsilon}
+ \frac{m^2(1{-}\Omega^2)^2}{16\pi^2(1{+}\Omega^2)^3}
\left(\int \! d^4u\ A_\mu(u)A_\mu(u)\right) \ln(\epsilon) \nonumber\\
 & +\frac{\Omega^2(1{-}\Omega^2)^2}{4\pi^2\theta^2(1{+}\Omega^2)^4}
\left(\int \!d^4u\ u^2A_\mu(u)A_\mu(u)\right) \ln(\epsilon)
\nonumber
\\
&- \frac{\Omega^4}{2\pi^2(1{+}\Omega^2)^4}\left(\int \!d^4u\ 
 (\widetilde{u}_\mu A_\mu(u))^2\right) \ln(\epsilon) \nonumber\\
& -\frac{(1{-}\Omega^2)^2(1{+}4\Omega^2{+}\Omega^4)}{
96\pi^2(1{+}\Omega^2)^4}
\left(\int \! d^4u\ A_\mu(u)\partial^2 A_\mu(u)\right) \ln(\epsilon) 
\nonumber\\
& -\frac{(1{-}\Omega^2)^4}{96\pi^2(1{+}\Omega^2)^4}
\left(\int \! d^4u\ (\partial_\mu A_\mu(u))^2\right)\ln(\epsilon)+ \dots
\\
\mathcal{T}_2'' &= -\frac{1}{16\pi^2(1{+}\Omega^2)}
\left(\int \!d^4u\ A_\mu(u)A_\mu(u)\right) \frac{1}{\epsilon}
- \frac{m^2}{16\pi^2(1{+}\Omega^2)}
\left(\int\! d^4u\ A_\mu(u)A_\mu(u)\right) \ln(\epsilon) \nonumber\\
 & -\frac{\Omega^2}{4\pi^2\theta^2(1{+}\Omega^2)^2}
\left(\int\! d^4u\ u^2A_\mu(u)A_\mu(u)\right) \ln(\epsilon) 
 \nonumber\\
  & + \frac{\Omega^2}{16\pi^2(1{+}\Omega^2)^2}
\left(\int d^4u\ A_\mu(u)\partial^2 A_\mu(u)\right) \ln(\epsilon) 
  + \dots
\end{align}
\end{subequations}

The computation of the 3-point function contributions can be
conveniently carried out by further using the following identity
\begin{equation}
\int \!d^4u\ \widetilde{u}_\mu A_\mu(u) (A_\nu\star A_\nu)(u) 
=\frac{1}{2}\int \! d^4u\ 
\left(\widetilde{u}_\mu A_\nu(u)\{A_\mu,A_\nu\}_\star(u) \; 
-i(\partial_\mu A_\nu(u))[A_\mu,A_\nu]_\star(u) +\frac{4}{\theta^2}\right).
\end{equation}
\begin{figure}[!htb]
  \centering
  \includegraphics[scale=1]{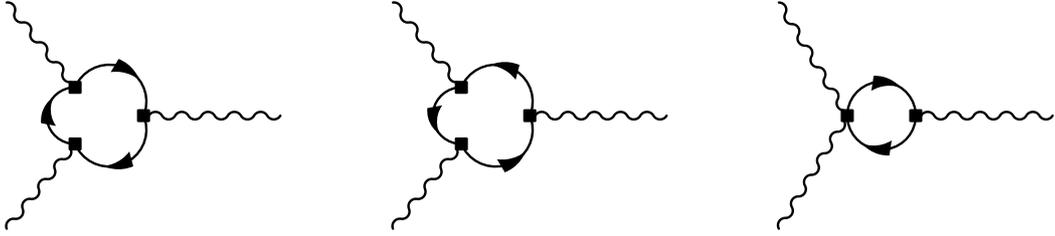}
  \caption[3-point]{\footnotesize{Relevant one-loop diagrams
      contributing to the 3-point function. Comments similar to those
      related to the figure 4 apply. The rightmost (resp.\ two
      leftmost) diagram(s) corresponds to the contribution
      $\mathcal{T}_3''$ (resp.\ $\mathcal{T}_3'$).}}
  \label{fig:3-point}
\end{figure}
The contributions corresponding to the
diagrams of figure \ref{fig:3-point} can then be expressed as
\begin{subequations}
\begin{align}
\mathcal{T}_3' &=
\frac{\Omega^2(1{-}\Omega^2)^2}{8\pi^2(1{+}\Omega^2)^4}
\left( \int \! d^4u\ \widetilde{u}_\mu A_\nu(u)\{A_\mu,A_\nu\}_\star(u) 
\right)\ln(\epsilon) \nonumber\\
 & +\frac{(1{-}\Omega^2)^2(1{+}4\Omega^2{+}\Omega^4)}{
48\pi^2(1{+}\Omega^2)^4}\left(\int \!d^4u\ (
(-i\partial_\mu A_\nu(u))[A_\mu,A_\nu]_\star(u)+\frac{4}{\theta^2} )\right)\ln(\epsilon)
+ \dots \\
\mathcal{T}_3'' &= -\frac{\Omega^2}{8\pi^2(1{+}\Omega^2)^2}
\left( \int \!d^4u\ (\widetilde{u}_\mu A_\nu(u)\{A_\mu,A_\nu\}_\star(u) 
+\frac{4}{\theta^2})\right)\ln(\epsilon) \nonumber\\
 & +\frac{i\Omega^2}{8\pi^2(1{+}\Omega^2)^2}
\left(\int \!d^4u\ (\partial_\mu A_\nu(u))[A_\mu,A_\nu]_\star(u)
\right)\ln(\epsilon)+ \dots
\end{align}
\end{subequations}
\begin{figure}[!htb]
  \centering
  \includegraphics[scale=1]{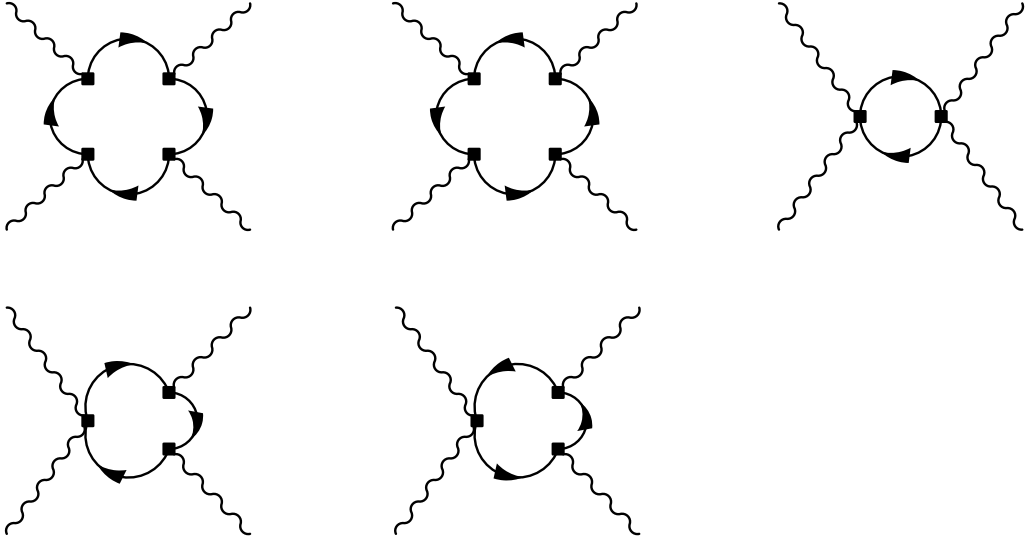}
  \caption[4-point]{\footnotesize{Relevant one-loop diagrams
      contributing to the 4-point function. Comments similar to those
      related to the figure 4 apply. Among the upper figures, the
      rightmost figure (resp.\ the two leftmost) diagram(s)
      corresponds to the contribution $\mathcal{T}_4'''$ (resp.\
      $\mathcal{T}_4'$). The lower diagrams correspond to
      $\mathcal{T}_4''$. }}
  \label{fig:4-point}
\end{figure}
In the same way, the 4-point contributions depicted on the figure
\ref{fig:4-point} are given by
\begin{subequations}
\begin{align}
\mathcal{T}_4' &= -\frac{(1{-}\Omega^2)^4}{96\pi^2(1{+}\Omega^2)^4} 
\left( \int \!d^4u\ \left( (A_\mu\star A_\nu (u))^2
+2(A_\mu\star A_\nu(u))^2\right) \right) \ln(\epsilon)+ \dots
\\
\mathcal{T}_4'' &= \frac{(1{-}\Omega^2)^2}{16\pi^2(1{+}\Omega^2)^2}
\left(\int \!d^4u\ (A_\mu\star A_\mu(u))^2 \right) \ln(\epsilon)+ \dots
\\
\mathcal{T}_4''' &= -\frac{1}{32\pi^2}\left(\int \! d^4u\ 
(A_\mu\star A_\mu(u))^2 \right) \ln(\epsilon)+ \dots
\end{align}
\end{subequations}

Finally, by collecting the various contributions given above, we find
that the effective action $\Gamma(A)$ can be written as
\begin{align}
\Gamma(A) &= \frac{\Omega^2}{4\pi^2(1{+}\Omega^2)^3}
\left(\int \!d^4u\ (\mathcal{A}_\mu\star\mathcal{A}_\mu 
-\frac{1}{4}\widetilde{u}^2) \right) 
\left(\frac{1}{\epsilon}+m^2\ln(\epsilon)\right) \nonumber
\\
& -\frac{(1{-}\Omega^2)^4}{192\pi^2(1{+}\Omega^2)^4} 
\left(\int \!d^4u\ F_{\mu\nu}\star F_{\mu\nu}\right) 
 \ln(\epsilon) \nonumber\\
 & +\frac{\Omega^4}{8\pi^2(1{+}\Omega^2)^4}
\left(\int \! d^4u\ (F_{\mu\nu}\star  F_{\mu\nu}
+\{\mathcal{A}_\mu, \mathcal{A}_\nu\}_\star^2 
-\frac{1}{4}(\widetilde{u}^2)^2)\right)\ln(\epsilon)+ \dots ,
 \label{eq:zegamma}
\end{align}
where $\mathcal{A}_\mu(u)= A_\mu(u)+{{1}\over{2}}\widetilde{u}_\mu$ and
$F_{\mu\nu}=\partial_\mu A_\nu-\partial_\nu
A_\mu-i[A_\mu,A_\nu]_\star$. To put the effective action into the form
\eqref{eq:zegamma}, it is convenient to use the following formulae
\begin{subequations}
\begin{align}
\int \!d^4x\ {\cal{A}}_\mu\star{\cal{A}}_\mu 
&=\int \!d^4x(\frac{1}{4}{\widetilde{x}}^2+{\widetilde{x}}_\mu A_\mu
+A_\mu A_\mu),
\\
\int \!d^4x\ F_{\mu\nu}\star F_{\mu\nu}
&= \int d^4x\big(\frac{16}{\theta^2}-2(A_\mu\partial^2A_\mu+(\partial_\mu A_\mu)^2)
-4i\partial_\mu A_\nu[A_\mu,A_\nu]_\star-[A_\mu,A_\nu]_\star^2 \big),
\\
\int \!d^4x\; \{{\cal{A}}_\mu,{\cal{A}}_\nu\}^2_\star
&=\int \!d^4x\; \big(\frac{1}{4}({\widetilde{x}}^2)^2
+2{\widetilde{x}}^2{\widetilde{x}}_\mu A_\mu
+4({\widetilde{x}}_\mu A_\mu)^2+2{\widetilde{x}}^2A_\mu A_\mu
\nonumber
\\
&+2(\partial_\mu A_\mu)^2
+4{\widetilde{x}}_\mu A_\nu\{A_\mu,A_\nu\}_\star
+\{A_\mu,A_\nu\}^2_\star\big).
\end{align}
\end{subequations}
The effective action \eqref{eq:zegamma} is one of the main results of
this paper. A somehow similar calculation can be performed when the
transformations correspond to those given in \eqref{eq:adjointde} and
the action \eqref{eq:adjcoupled}. It turns out that the non-planar
graphs are UV finite so that the corresponding effective action
$\Gamma_{adj}(A)$ satisfies
\begin{equation}
\Gamma_{adj}(A)=2\, \Gamma(A).
\end{equation}

\section{Discussion.}

Let us summarise and discuss the results we obtained. In this paper,
we considered the involutive unital Moyal algebra ${\cal{M}}$ in 4
space dimensions, as described in section 2, and focused on
noncommutative field theories defined on ${\cal{M}}$ viewed as a
(hermitian) module over itself. We started from a renormalisable
scalar field theory which can be viewed as the extension to complex
valued fields $\phi$ of the renormalisable noncommutative $\varphi^4$
with harmonic term studied in \cite{Grosse:2004yu,Gurau:2005gd}. By
further applying a minimal coupling prescription, that we discussed in
section 2, this action is coupled to an external gauge potential and
gives rise to a gauge-invariant action $S(\phi,A)$, the point of
departure for the computation of the effective action $\Gamma(A)$. The
whole analysis is based on the usual algebraic construction of
connections relevant to a noncommutative framework. As presented in
section 2, the modules of the algebra play the role of the set of
sections of vector bundles of ordinary geometry while the
noncommutative analogue of gauge transformations are naturally
associated with the automorphisms of (hermitian) modules. The fact
that ${\cal{M}}$ involves only inner derivations implies the existence
of a gauge-invariant connection which is further used as a reference
connection. It plays a special role in the minimal coupling
prescription and permits one to relate the so-called covariant
coordinates \cite{Douglas:2001ba} to a tensorial form built from the
difference of two connections. We also pointed out that scalar fields
which transform under the adjoint representation of the gauge group do
not fit into the above algebraic framework, because noncommutative
gauge transformations are automorphisms of modules while ``adjoint
transformations'' are automorphisms of the algebra.
\par

We have computed at the one-loop order the effective action
$\Gamma(A)$ given in \eqref{eq:zegamma}, obtained by integrating over
the scalar field $\phi$, for any value of the harmonic oscillator
parameter $\Omega \in [0,1]$ in $S(\phi,A)$. Details of the
calculation are collected in the section 3. We find that the effective
action involves, beyond the usual expected Yang-Mills contribution
\mbox{$\sim \int d^4x\ F_{\mu\nu}\star F_{\mu\nu}$}, additional terms
of quadratic and quartic order in ${\cal{A}}_\mu$
\eqref{eq:arondvrai}, $\sim \int d^4x\
{\cal{A}}_\mu\star{\cal{A}}_\mu$ and $\sim \int d^4x\
\{{\cal{A}}_\mu,{\cal{A}}_\nu\}_\star^2$. These additional terms are
gauge invariant thanks to the gauge transformation of ${\cal{A}}_\mu$
\eqref{eq:arondfaux}. The quadratic term involves a mass term for the
gauge potential $A_\mu$ (while such a bare mass term for a gauge
potential is forbidden by gauge invariance in commutative Yang-Mills
theories). We further notice that the presence of a quartic term $\sim
\int d^4x\ \{{\cal{A}}_\mu,{\cal{A}}_\nu\}_\star^2$ accompanying the
standard Yang-Mills term is reminiscent to the occurrence of a
(covariance under a) Langmann-Szabo duality \cite{Langmann:2002cc}.
Basically, Langmann-Szabo duality is generated through the exchange
$i\partial_\mu \leftrightarrows {\widetilde{x}}_\mu$ which, upon using
\eqref{eq:innerder} and $\{{\widetilde{x}}_\mu,f\}_\star =
2{\widetilde{x}}_\mu f$, can be expressed as $[\xi_\mu,.]_\star
\leftrightarrows \{\xi_\mu,.\}_\star$.  This, combined with
\eqref{eq:f-arond}, therefore suggests that some covariance under
Langmann-Szabo duality would show up whenever both commutators and
anti-commutators are involved in the action. By the way, at the
special value $\Omega = 1$ where the scalar model considered in
\cite{Langmann:2002cc} is duality-invariant, the effective action
\eqref{eq:zegamma} is fully symmetric under the exchange
$[{\cal{A}}_\mu,{\cal{A}}_\nu]_\star \leftrightarrows
\{{\cal{A}}_\mu,{\cal{A}}_\nu\}_\star$.\par

Recently, a calculation based on the machinery of Duhamel expansions
of the (one-loop) action for the effective gauge theory stemming from
a (real-valued) scalar theory with harmonic term has been carried in
\cite{Grosse:2006hh}. The scalar theory considered in
\cite{Grosse:2006hh} was somehow similar to the one described by the
action \eqref{eq:adjcoupled} together with transformations as those
given in \eqref{eq:adjointde}. The analysis was performed within the
matrix base so that, due to the complexity of the calculations, in
four dimensions only the case $\Omega = 1$ was considered. Our result
for the effective action $\Gamma_{adj}(A)$ agrees globally with the
one given in \cite{Grosse:2006hh}, up to unessential numerical
factors.  Notice that the calculations are easier within the $x$-space
formalism even when $\Omega \neq 1$. \par

At this point, one important comment relative to \eqref{eq:zegamma} is
in order. The fact that the tadpole is non-vanishing (see \eqref{eq:tasdepaul}) is a rather unusual feature for a Yang-Mills
type theory. This non-vanishing implies automatically the occurrence
of the mass-type term $\int d^4x\ {\cal{A}}_\mu\star{\cal{A}}_\mu$ as
well as the quartic term $\int
d^4x\ \{{\cal{A}}_\mu,{\cal{A}}_\nu\}_\star^2$. Keeping this in mind
together with the expected impact of the Langmann-Szabo duality on
renormalisability, it is tempting to conjecture that the following
class of actions
\begin{align}
S=\int d^4x \Big(\frac{\alpha}{4g^2}F_{\mu\nu}\star F_{\mu\nu}
+\frac{\Omega^\prime}{4g^2}\{{\cal{A}}_\mu,{\cal{A}}_\nu\}^2_\star
+\frac{\kappa}{2} {\cal{A}}_\mu\star{\cal{A}}_\mu\Big) \label{eq:decadix}
\end{align}
involves suitable candidates for renormalisable actions for gauge
theory defined on Moyal spaces. Recall that the naive action for
a Yang-Mills theory on the Moyal space, $\sim \int d^4x\
F_{\mu\nu}\star F_{\mu\nu}$, exhibits UV/IR mixing
\cite{Hayakawa:1999yt,Matusis:2000jf}, making its renormalisability
quite problematic. In \eqref{eq:decadix}, the second term built from
the anticommutator may be viewed as the ``gauge counterpart'' of the
harmonic term ensuring the renormalisability of the $\varphi^4$ theory
investigated in \cite{Grosse:2004yu}, while $\alpha$, $\Omega^\prime$
and $\kappa$ are real parameters and $g$ denotes some coupling
constant. According to the above discussion, the presence of the
quadratic and quartic terms in ${\cal{A}}_\mu$ in \eqref{eq:decadix}
will be reflected in a non-vanishing vacuum expectation value for
$A_\mu$. The consequences of a possible occurrence of this non-trivial
vacuum remain to be understood and properly controlled in view of a
further gauge-fixing of a (classical) gauge action stemming from
\eqref{eq:decadix} combined with a convenient regularisation scheme
(that could be obtained by some adaptation of
\cite{Rivasseau:2007qx}). We will come back to these points in a
forthcoming publication.\par

\vskip 1 true cm 

\noindent
{\bf{Acknowledgements}}: We are grateful to M.
Dubois-Violette, H. Grosse, R. Gurau, J. Magnen, T. Masson, V.
Rivasseau and F. Vignes-Tourneret for valuable discussions at various
stages of this work. This work has been supported by ANR grant
NT05-3-43374 "GenoPhy". \par

\end{document}